# Thermal tunability in terahertz metamaterials fabricated on strontium titanate single crystal substrates

**Ranjan Singh,**[a)] **Abul K. Azad, Q. X. Jia, Antoinette J. Taylor, and Hou-Tong Chen**[b)]

*Center for Integrated Nanotechnologies, Los Alamos National Laboratory, Los Alamos, New Mexico 87545, USA*

E-mail: a) ranjan@lanl.gov; b) chenht@lanl.gov

**Abstract**

We report an experimental demonstration of thermal tuning of resonance frequency in a planar terahertz metamaterial consisting of a gold split-ring resonator array fabricated on a bulk single crystal strontium titanate ($SrTiO_3$) substrate. Cooling the metamaterial starting from 409 K down to 150 K causes about 50% shift in resonance frequency as compare to its room temperature resonance, and there is very little variation in resonance strength. The resonance shift is due to the temperature-dependent refractive index (or the dielectric constant) of the strontium titanate. The experiment opens up avenues for designing tunable terahertz devices by exploiting the temperature sensitive characteristic of high dielectric constant substrates and complex metal oxide materials.

Metamaterials have attracted great attention of physicists and engineers as they can be used to fabricate devices with novel functionalities over a large electromagnetic spectral domain. Split-ring resonators (SRRs) [1] form the most important component of metamaterials, resonantly coupling with the incident electric and/or magnetic fields and possessing the key characteristic of being much smaller in size than their resonance wavelength. Therefore, SRRs have tremendous potential to be considered as fundamental elements suitable for miniaturization of planar microwave, terahertz (THz) and optical devices such as notch filters, sensors, and antennas [2-4]. However, in order to fully exploit the exotic properties they offer and to extend functionality of metamaterial devices, we need to have active or dynamic control over their resonances [5-8], which is of particular significance in the THz frequency band due to the severe lack of THz materials provided by nature. Recently, frequency tunability of THz metamaterials has been explored, for instance by incorporating semiconductors [8,9], insulator-to-metal transition films [10-11], and superconducting metamaterials [12,13], which are accomplished by using external stimuli such as thermal, optical excitation, and electric and magnetic fields. In all these cases the frequency tuning of metamaterial resonance is accompanied with a large variation of resonance strength, which is undesirable and caused by the damping from the substrate or metamaterial elements.

In this letter, we report an experimental demonstration of thermally tuning the frequency of the fundamental metamaterial resonance by up to 50%, as compared to its resonance frequency at room temperature. This is accomplished by the temperature-dependent refractive index (or the dielectric constant) of the bulk single crystal strontium titanate ($SrTiO_3$ or STO) substrate, where the metamaterial resonance strength does not suffer significant variation. STO has attracted growing attention for the next generation oxide electronics since it exhibits unique electrical properties [14], which can be easily engineered by appropriately substituting dopants and controlled by temperature or electrical bias. Due to its crystalline structure and chemistry compatibility, commercially available single crystal STO is being widely used to grow heteroepitaxial films of many perovskite oxides such as high temperature cuprate superconductors and manganates. STO is also being seen as a prospective material for use in the radio and microwave frequency devices due to its high and tunable dielectric constant and low loss. In this experiment, we demonstrate for the first time that bulk single crystal STO can be used for designing frequency agile THz metamaterials by tuning its refractive index with temperature.

The metamaterial was fabricated as a planar square array of subwavelength 200-nm-thick gold electric SRRs [2] on a 533-μm-thick single crystal (100) oriented STO substrate, as shown in Fig. 1, using conventional photolithography, electron beam metal film deposition, and lift-off process. The geometrical dimensions of the SRR unit cell are shown in the inset of Fig. 1. Under normal incidence, the metamaterial was characterized as a function of temperature using a THz time-domain spectroscopy (TDS) system [15] incorporated with a continuous flow liquid helium cryostat. It utilizes a 1 kHz regeneratively amplified Ti:sapphire near-infrared femtosecond laser for THz generation and detection using ZnTe crystals via optical rectification and the electro-optic effect. By Fourier transforming the pulsed THz time-domain signals transmitted through the metamaterial sample and a bare STO as the reference, the normalized transmission amplitude spectra are extracted as their ratio, $t(\omega) = |E_s(\omega)/E_r(\omega)|$, where $E_s(\omega)$ and $E_r(\omega)$ are for the sample and reference, respectively.

Figure 2(a) shows the measured time-domain waveforms of the transmitted THz pulses through the metamaterial sample and the reference at different temperatures, where the few-cycle oscillating tails indicate a resonant response of the metamaterial. The THz pulse is delayed by 4.5 ps in time as the sample is cooled down from 409 K to room temperature (295 K). It is further delayed by 7.3 ps and 7.7 ps respectively as the temperature decreases from 295 K to 200 K and from 200 K to 150 K. The total change of time delay between 409K and 150 K is about 19.6 ps, which can be used to estimate the change of refractive index approximately $\Delta n \approx 11$ of the STO. This is in good agreement with the measured temperature-dependent refractive index of a bare STO shown in Fig. 3(a), which is obtained at different temperatures from THz-TDS measurements [16] and is 14.8 at 409 K and 28.8 at 150 K, giving a change of refractive index $\Delta n = 14$. It is worth noting that the increasing high refractive index of STO also results in higher insertion loss in THz transmission at lower temperatures, as indicated by smaller magnitude of overall transmitted THz pulses for both the metamaterial and the reference. This, however, may be alleviated for example by using metamaterial antireflection coatings [17].

The normalized THz transmission amplitude spectra are shown in Fig. 2(b) at the corresponding temperatures. At all these temperatures, the metamaterial exhibits a strong transmission dip due to the resonant excitation of circulating currents by the incident THz electric field [2]. The resonance experiences a gradual red shift from 0.223 THz at 409 K to 0.128 THz at 150 K. The total frequency shift in resonance is 95 GHz, which is about 50% of the resonance frequency at room temperature (0.194 THz). This is due to the temperature dependent refractive index of the STO, which is shown in Fig. 3(a) and reveals an increasing refractive index with decreasing temperature. Since the fundamental resonance frequency of SRRs is approximately given by $\omega_0 = (LC)^{-1/2}$, where $L$ is the loop inductance and $C$ is the gap capacitance, the latter is directly proportional to the dielectric constant $\varepsilon = n^2$ (when the loss is negligible) of the STO, i.e. $\omega_0 \sim 1/n$. This is verified by the plots shown in Fig. 3(b), where excellent agreement is achieved in the temperature dependence of the resonance frequency $\omega_0$ and the inverse of refractive index $1/n$ taken at the corresponding resonance frequencies.

In contrast to other frequency tuning methods [8-13], here the resonance strength shows very little variation as suggested by the similar values of transmission minimum as well as quality factor. In our experiments the loss tangent of the STO, which is about 0.02 at 0.2 THz, largely does not change with temperature. Further tuning the metamaterial resonance with lower temperatures is prevented due to the following two reasons. First, the resonance is expected to shift further to lower frequencies, beyond the measurement capability of our THz-TDS system. Second, the refractive index of the substrate continues to increase rapidly [18] due to the phase transition so that the transmission signal becomes too small, as indicated by the time-domain THz signal shown in Fig. 2(a) with decreasing temperatures. This might be overcome by using STO thin films and only loading STO at the critical regions at SRRs, which has been demonstrated in the microwave frequency range where an electrical tuning of metamaterial resonance was accomplished [19]. It should be also noted that the compatibility of STO (both bulk and thin film) with other complex metal oxides such as high temperature superconducting metamaterials [12] can open multi-functional THz metamaterial devices where the resonance can be also controlled through various approaches.

In summary, we have measured in the THz frequency range the resonant behavior of a planar gold SRR array fabricated on single crystal bulk STO when varying the temperature. We observed 50% resonance frequency tuning without significant variation of resonance strength. The resonance tuning occurs due to the temperature-dependent refractive index (or the dielectric constant) of the STO. Such a thermal tuning of metamaterial resonance using STO and ferroelectric materials will enable the integration of metamaterials with other complex metal oxides and resonance tuning approaches to realize multi-functional THz metamaterial devices.

We acknowledge support from the Los Alamos National Laboratory LDRD Program. This work was performed, in part, at the Center for Integrated Nanotechnologies, a US Department of Energy, Office of Basic Energy Sciences Nanoscale Science Research Centre operated jointly by Los Alamos and Sandia National Laboratories. Los Alamos National Laboratory, an affirmative action/equal opportunity employer, is operated by Los Alamos National Security, LLC, for the National Nuclear Security Administration of the US Department of Energy under contract DE-AC52-06NA25396.

**References**

1. J. B. Pendry, A. J. Holden, D. J. Robbins, and W. J. Stewart, IEEE Trans. Microwave Theory Tech. **47**, 2075 (1999).

2. H.-T. Chen, J. F. O'Hara, A. J. Taylor, R. D. Averitt, C. Highstrete, M. Lee, and W. J. Padilla, Opt. Express **15**, 1084 (2007).

3. J. F. O'Hara, R. Singh, I. Brener, E. Smirnova, J. Han, A. J. Taylor, and W. Zhang, Opt. Express **16**, 1786 (2008).

4. K. B. Alici and E. Ozbay, J. Appl. Phys. **101**, 083104 (2007).

5. W. J. Padilla, A. J. Taylor, C. Highstrete, M. Lee, and R. D. Averitt, Phys. Rev. Lett. **96**, 107401 (2006).

6. H.-T. Chen, W. J. Padilla, J. M. O. Zide, A. C. Gossard, A. J. Taylor, and R. D. Averitt, Nature **444**, 597 (2006).

7. H.-T. Chen, W. J. Padilla, M. J. Cich, A. K. Azad, R. D. Averitt, and A. J. Taylor, Nat. Photon. **3**, 148 (2009).

8. H.-T. Chen, J. F. O'Hara, A. K. Azad, A. J. Taylor, R. D. Averitt, D. B. Shrekenhamer, and W. J. Padilla, Nat. Photon. **2**, 295 (2008).

9. N.-H. Shen, M. Kafesaki, T. Koschny, L. Zhang, E. N. Economou, and C. M. Soukoulis, Phys. Rev. B **79**, 161102(R) (2009).

10. T. Driscoll, S. Palit, M. M. Qazilbash, M. Brehm, F. Keilmann, B.-G. Chae, S.-J. Yun, H.-T. Kim, S. Y. Cho, N. M. Jokerst, D. R. Smith, and D. N. Basov, Appl. Phys. Lett. **93**, 024101 (2008).

11. T. Driscoll, H.-T. Kim, B.-G. Chae, B.-J. Kim, Y.-W. Lee, N. M. Jokerst, S. Palit, D. R. Smith, M. Di Ventra, and D. N. Basov, Science **325**, 1518 (2009).

12. H.-T. Chen, H. Yang, R. Singh, J. F. O'Hara, A. K. Azad, Q. X. Jia, and A. J. Taylor, Phys. Rev. Lett. **105**, 247402 (2010).

13. B. B. Jin, C. H. Zhang, S. Engelbrecht, A. Pimenov, J. B. Wu, Q.Y. Xu, C. H. Cao, J. Chen, W.W. Xu, L. Kang, and P. H. Wu, Opt. Express **18**, 17504 (2010).

14. M. J. Dalberth, R. E. Stauber, J. C. Price, C. T. Rogers, and D. Galt, Appl. Phys. Lett. **72**, 507 (1998).

15. R. D. Averitt and A. J. Taylor, J. Phys. Condens. Matter **14**, R1357 (2002).

16. W. Zhang, A. K. Azad, and D. Grischkowsky, Appl. Phys. Lett. **82**, 2841 (2003).

17. H.-T. Chen, J. Zhou, J. F. O'Hara, F. Chen, A. K. Azad, and A. J. Taylor, Phys. Rev. Lett **105**, 073901 (2010).

18. H. Nemec, P. Kuzel, L. Duvillaret, A. Pashkin, M. Dressel, and M. T. Sebastian, Opt. Lett. **30**, 549 (2005).

19. T. H. Hand and S. A. Cummer, J. Appl. Phys. **103**, 066105 (2008).

**Figure Captions**

Fig. 1 Microscopic image of a metamaterial consisting of a 200-nm-thick gold electric split-ring resonator array fabricated on a strontium titanate substrate. The inset is an individual unit cell with dimensions shown.

Fig. 2 (a) THz pulses measured in the time-domain after transmitting through the metamaterial sample (red solid) or a bare strontium titanate substrate as the reference (blue dashed) at various temperatures. The curves at different temperatures have been vertically translated for clarity. (b) The corresponding normalized THz transmission amplitude spectra through the metamaterial.

Fig. 3 (a) Experimentally measured refractive index $n$ of the strontium titanate substrate by THz time-domain spectroscopy at various temperatures. (b) A comparison between the resonance frequency $\omega_0$ and the inverse of refractive index indicating $\omega_0 \sim 1/n$. The values of $n$ have been taken from (a) at the corresponding metamaterial resonance frequencies.

**Fig. 1**

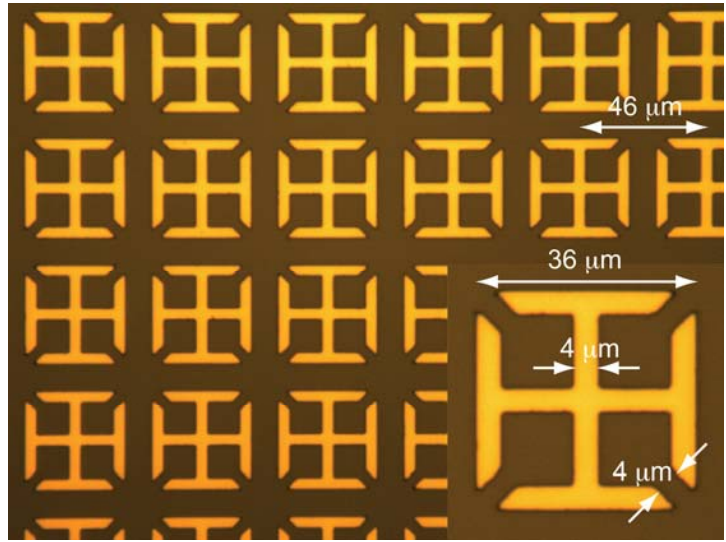

**Fig. 2**

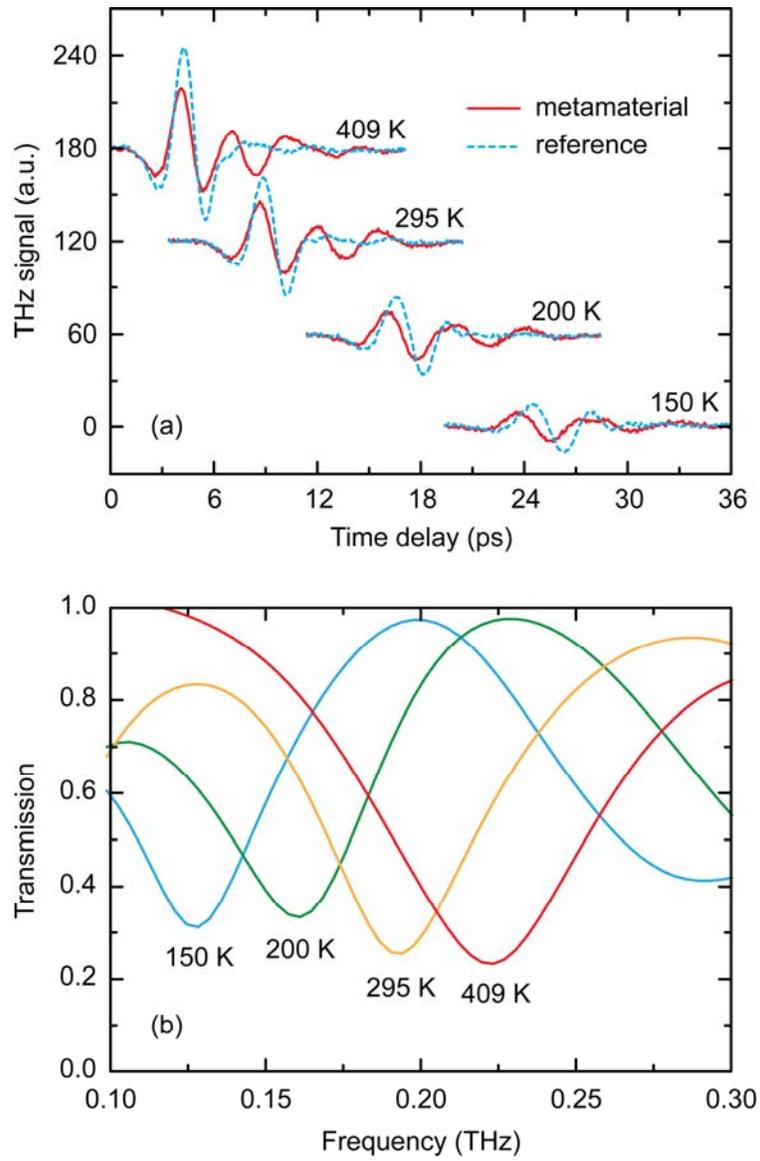

**Fig. 3**

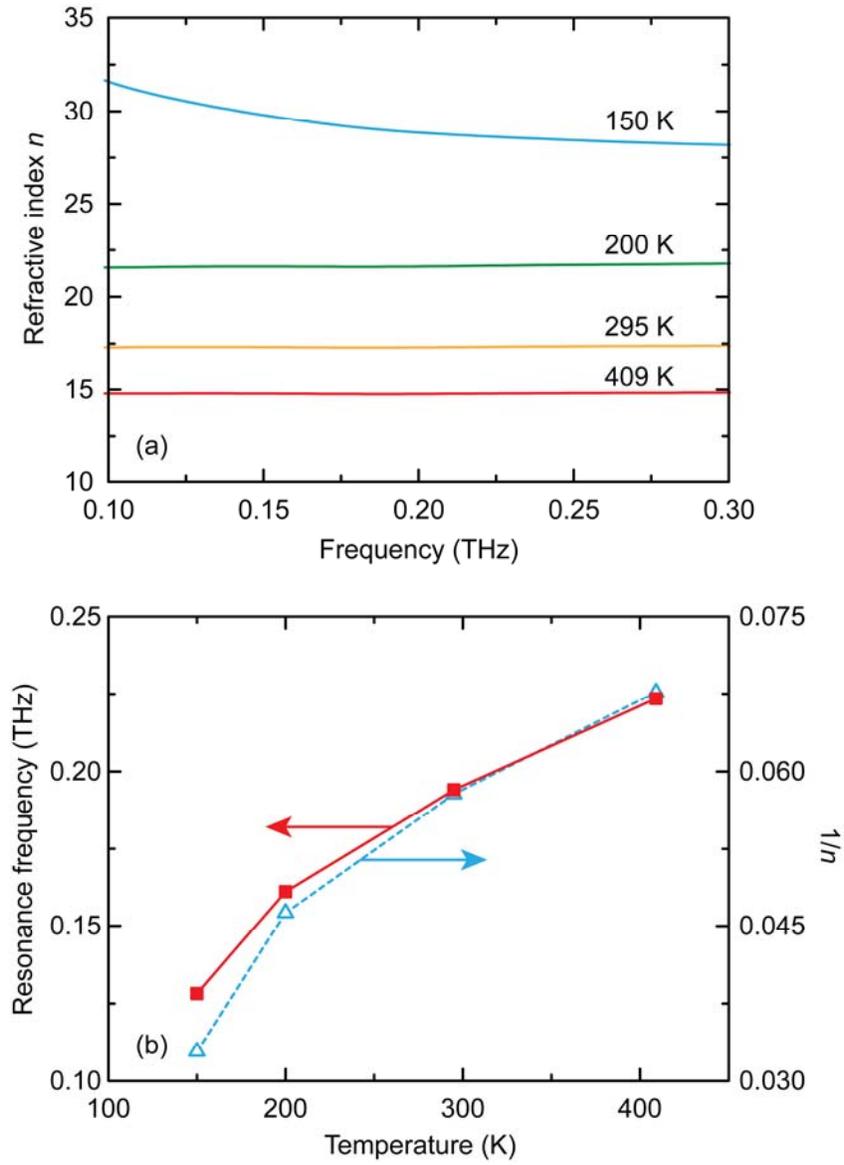